# High pressure high temperature (HPHT) synthesis and magnetism of $MSr_2Y_{1.5}Ce_{0.5}Cu_2O_{10}$ compounds with M = Ru, Mn, and Cr

V.P.S. Awana[1,2,*], S. Balamurugan[2], H. Kishan[1], and E. Takayama-Muromachi[2]

[1]National Physical Laboratory, Dr. K.S. Krishnan Marg, New Delhi-110012, India
[2]Superconducting Materials Center, NIMS, 1-1 Namiki, Tsukuba, Ibaraki, 305-0044, Japan

We report synthesis and magnetization of $MSr_2Y_{1.5}Ce_{0.5}Cu_2O_{10}$ (M/Y-1222) compounds with M = Ru, Mn and Cr by HPHT. They are synthesized at high pressure of 6 GPa in optimized temperatures of up to 1450 $^0$C, for Ru, Mn and 1550 $^0$C in case of Cr based Y/1222, for a time duration of three hours for all the samples. All the samples crystallized with tetragonal structure in *I*4/*mmm* space group. Few minor impurities are also seen in case of Mn,Cr/Y-1222 samples. DC magnetic susceptibility measurements exhibited ferromagnetic–like transition for all the three compounds at low temperatures. At 5 K, the magnetization $M(H)$ experiments showed clear ferromagnetic like hysteresis loops for Ru and Mn samples with sizeable returning moment ($M_r$) and coercive field ($H_c$). In case of Cr/Y-1222 system though, ferromagnetism is visible below around 45 K in susceptibility measurements, the characteristic $M_r$ and $H_c$ are not seen. To the best of our knowledge the Mn,Cr/Y-1222 phases are synthesized for the first time and the quality of our earlier reported Ru/Y-1222 is improved.

# I. INTRODUCTION

Superconductivity and magnetism are not supposed to coexist in a single thermodynamic phase, even then there are some systems, which show the coexistence of the two [1-3]. In this regards rutheno-cuprate ferromagnetic superconductors had attracted a lot of attention [3-5]. In case of rutheno-cuprates, magneto-superconductivity was first discovered in $RuSr_2(Sm,Gd,Eu)_{1.5}Ce_{0.5}Cu_2O_{10}$ (Ru-1222) [3], and later in $RuSr_2(Gd,Eu,Sm)Cu_2O_8$ (Ru-1212) [4]. The scientific opinion is yet divided on the issue of genuine coexistence of magnetism and superconductivity in these systems [5,6]. Most of the experimental studies on the ruthenocuprates are confined to $RuSr_2Gd_{1.5}Ce_{0.5}Cu_2O_{10}$ (Gd/Ru-1222) and $RuSr_2GdCu_2O_8$ (Gd/Ru-1212). The presence of high paramagnetic moment of $Gd^{3+}$ ($8\mu_B$) in these compounds complicates the interpretation of coexisting magnetism and superconductivity in them. Further, Gd, Sm and Eu absorb the thermal neutrons, and hence it was difficult to collect good data from neutron powder diffraction for the magnetic structure studies. All this warranted the need for synthesis of Y/Ru-1212 and Y/Ru-1222. First the Y/Ru-1212 was synthesized by high-pressure-high-temperature (HPHT) independently by couple of research groups including us [7-9], later in year 2003 we reported the synthesis of Y/Ru-1222 by HPHT process [10].

Interestingly enough, when Ru in Ru-1212/Ru-1222 is completely replaced by other 3d metals viz. Fe or Co the phenomenon of coexisting superconductivity and magnetism is not observed [11-13]. In this regard the trials for other magnetic-ion replacements at Ru site are obvious. The Ru could be replaced fully by Fe and Co in either Ru-1212 or 1222 by normal synthesis routes itself [11-13], but not by some other 3d metals viz., Mn and Cr. We tried Mn and Cr replacement at Ru site in Ru-1222 by HPHT process. It is known that HPHT often leads to the formation of compounds, which cannot be formed by any other normal technique. Another advantage of HPHT is that one can nearly fix the oxygen content of the compound, as whole process of synthesis takes place in a closed cell environment.

By following our previously reported route of the HPHT synthesis of Y/Ru-1222 [10], we not only have improved the quality of Y/Ru-1222 but could synthesize successfully for the first time the Mn/Y-1222 and Cr/Y-1222 compounds. Besides the novel synthesis of Ru,Mn,Cr/Y-1222 phases we also report their magnetization and show that these systems exhibit low temperature weak ferromagnetism. Though presently studied compounds are not magneto-superconducting they could yet open up a new front to look for various magneto-superconducting cuprates other than the fascinating rutheno-cuprates, which are more or less studied widely over a span of more than half a decade by now.

## II. EXPERIMENTAL DETAILS

Samples of composition $MSr_2Y_{1.5}Ce_{0.5}Cu_2O_{10}$ with M = Ru, Mn and Cr were synthesized through a HPHT solid-state reaction route. For the HPHT synthesis and to fix the oxygen at 10.0 level, the molar ratio used were: $(RuO_2) + 2(SrO_2) + 3/4(Y_2O_3) + 1/2(CeO_2) + 3/4(Cu_2O) + 1/2(Cu)$ resulting in $RuSr_2Y_{1.5}Ce_{0.5}Cu_2O_{10}$, for Ru/Y-1222, $(MnO_2) + 2(SrO_2) + 3/4(Y_2O_3) + 1/2(CeO_2) + 3/4(Cu_2O) + 1/2(Cu)$ resulting in $MnSr_2Y_{1.5}Ce_{0.5}Cu_2O_{10}$, for Mn/Y-1222 and $1/2(Cr_2O_3) + 2(SrO_2) + 3/4(Y_2O_3) + 1/2(CeO_2) + 5/4(CuO) + 3/4(Cu)$ resulting in $CrSr_2Y_{1.5}Ce_{0.5}Cu_2O_{10}$, for Cr/Y-1222. The materials were mixed in an agate mortar. Later around 300 mg of the mixture was sealed in a gold capsule and allowed to react in a flat-belt-type-high-pressure apparatus at 6GPa and 1200 $^0$C – 1550 $^0$C depending upon M (Ru, Mn or Cr) for 3 hours. After several experimentations of synthesis temperature runs, the optimized temperatures of synthesis for phase formation at 6 GPa and 3 hours were found to be 1450 $^0$C, for Ru, Mn and 1550 $^0$C in case of Cr based Y-1222 compounds. Nearly no change was observed in the weight of synthesized samples, indicating towards their fixed nominal oxygen content. However this contention is not precisely correct, and hence exact value of oxygen content of these samples is yet warranted through thermogravimetric (TG) analysis or chemical titration. X-ray powder diffraction patterns were obtained by a diffractometer with Cu K$_\alpha$ radiation. DC susceptibility data were collected by a SQUID magnetometer (Quantum Design, MPMS).

## III. RESULTS AND DISCUSSION

The X-ray diffraction patterns of the M/Y-1222 compunds with M = Ru, Mn and Cr are depicted in Fig. 1. The main intensity lines of the patterns i.e. the low angle high intensity with *hkl* [107] and high angle low intensity of *hkl* [110], corresponding to tetragonal structure with *I*4/*mmm* space group, are visible for all the three samples. Further, these main intensity peaks neither fit to the M/Y-1212 ($MSr_2YCu_2O_8$) nor to M-211O$_6$ ($Sr_2MYO_6$). The characteristic M-211O$_6$ main structural peak at $2\theta \equiv 30^0$ is also not seen in any of the three studied samples. In case of various M-12S2 compounds with S = 1, 2, or 3, one can identify them separately. In particular the characteristic low angle (001) peak appears for tetrgagonal *P*4/*mmm* sapace group M-1212 and M-1232 respectively at around $d$ = 11.5 Å and 16 Å, reflecting their *c*-parameter values. In case of M-1222 the space group is changed to *I*4/*mmm*, with doubled *c*-parameter hence instead of (001), the characteristic (002) peak appears at close to 14 Å. Intergrowth of M-1212, 1222 and 1232 phases give rise to multiple peaks in the range of d = 11 Å to 16

Å. This is demonstrated by some of us in the case of M = Co in Ref. 14, where the XRD patterns of all the three members are compared with each other. In the present case we get only one peak close to d = 14 Å, without any intergrowth, please see inset in Fig.1 for the case of Cr/Y-1222. The presence of only characteristic low angle peak of 1222 without the intergrowth of either 1212 or 1232 phases, excludes the possibility of 1212 phase in our sample. At this stage we believe, our compounds are crystallized mostly in tetragonal structure with *I*4/*mmm* space group.

As far as Ru/Y-1222 in concerned its complete XRD pattern could be indexed very well for tetragonal structure with *I*4/*mmm* space group. In fact the x-ray quality, in relation to the presence of minor impurities, of our presently studied Ru/Y-1222 compound is better than the one we reported earlier in Ref. 10. The improved phase purity of the compound is reflected in magnetization measurements, which we will discuss later. In case of Mn and Cr based Y/1222 compounds though some small intensity x-ray diffraction lines can be indexed due to minor impurities, the most others belonged to the tetragonal structure with *I*4/*mmm* space group same as for Ru/Y-1222. There also seems to appear some extra superstructure lines for Mn and Cr based Y/1222 compounds. This is possible in case of Mn and Cr based Y/1222 compounds due to the presence of their varying mixed valences and as a result the possible ordering of $MO_6$/$MO_4$ in both *a* and *c* directions. This situation is different for Ru/Y-1222 compound where due to presence of mostly $Ru^{5+}$, the $RuO_6$ octahedras are uniformly formed in the structure, with fewer superstructures. Some of us have earlier reported the detailed micro-structural studies on some of the M-12S2 compounds, viz., Ru and Co please see Ref. 15 and 16. Our presently investigated samples warrant micro-structural studies, which will be taken up soon and the results will be reported elsewhere. What we can safely conclude from our present XRD results in Fig. 1 is that (a): the quality of previously HPHT synthesized Ru/Y-1222 is improved significantly, in terms of its phase purity, by increasing the heating temperature and (b) the Mn and Cr based Y/1222 compounds are synthesized for the first time in near single phase, though further optimization is still needed. As far as lattice parameters are concerned there is not much difference between presently studied M-1222 compounds. For Ru/Y-1222 the same are *a* = 3.81 Å and *c* = 28.43 Å. Also as we mentioned above, all reflections for Mn and Cr samples could not be accounted within the same structure as for Ru/Y-1222, hence we do not find it fit to compare their lattice parameters at this stage, before exactly knowing the possible superstructures for Mn and Cr based Y/1222 compounds.

The main panel of Fig. 2 depicts the DC magnetic susceptibility ($\chi$) versus temperature (*T*) plot for Ru/Y-1222 compound in an applied field of 10 Oe. Both zero-field-cooled (ZFC) and field-cooled (FC) parts of $\chi$ are shown in the main panel. The general behavior of the $\chi(T)$ plot in terms of magnetic ordering temperature ($T_{mag.}$), seen as branching of ZFC and FC at around 130 K and a cusp in ZFC at 90 K is very similar to that as reported earlier [10] for our HPHT synthesized Y/Ru-1222 sample. The magnetization $M(H)$ plots of the compound at 5 K in low fields of up to 3000 Oe and higher field of up to 5 Tesla (T) are shown respectively in inset –II and inset – III of Fig. 2. Clear ferromagnetic loop like opening of the $M(H)$ plot is visible in both the insets – II and – III. The coercive field ($H_c$) is around 250 Oe and returning moment ($M_r$) is close to 0.32 $\mu_B$. These values are comparable to that as reported in Ref. 10.

One difference is that though diamagnetism is observed in $\chi(T)$ during ZFC for inset I sample, the same is not seen in the present case. In fact the earlier sample (Inset I) was also shown to be non-superconducting with subsequent magnetization [17] and resistivity [Ref. 10, note added in proofs] measurements. The negative signal in ZFC process may appear due to small-trapped negative field in the SQUID even for a non-superconducting compound. This fact need to be taken care of more specifically in relation to magneto-superconducting compounds, where the FC signal can not be diamagnetic even in the presence of bulk superconductivity in the system [9]. In present case we have *nullified* the trapped negative moment in the SQUID before the ZFC process measurements and was checked by measuring the straw without sample. Furthermore the sample does not show any superconducting transition in the resistivity measurements. Hence we can conclude that present sample of Ru/Y-1222 is not superconducting. Also to best of our knowledge the Y/Ru-1222 compounds yet reported [10, 17] are all magnetic at low temperatures, but not superconducting. The situation is though different for Gd,Sm,Eu/Ru-1222 compounds which are magneto-superconducting [3,5,6,16,18].

An important point about the magnetism of Ru/Y-1222 is about the various magnetic transitions of Ru in this compound. This is discussed by some of us in our earlier articles [10, 17]. Though the $M_r$ decreases monotonically with increasing temperature until $T_{mag}$, in the same way as being expected for a normal ferromagnetic type transition, $H_c$ exhibited an increase at intermediate temperatures with further decrease and becoming zero above $T_{mag}$. To elucidate this point on our better quality sample, we measured its $M(H)$ at low fields. Nice ferromagnetic like $M(H)$ plots are obtained, please see Fig. 3. Interestingly enough, it was seen that though $M_r$ decreases monotonically with increase in $T$, the $H_c$ becomes nearly zero at 50-60 K (loop opening is nearly diminished) followed by further increases of the same with clear opening of the loop at above 90 K and finally dropping again to zero above $T_{mag}$, please see $H_c(T)$ plot in inset of Fig. 3. It seems some short of re-alignment of Ru spins is taking place at around 80-100 K, which is close to the cusp temperature of ZFC susceptibility. We do not need to discuss this at length as it is already elaborated earlier by some of us [10, 17, 18]. The point is that the intriguing behavior of $H_c$ is also seen in presently studied cleaner Ru/Y-1222 sample.

The magnetization results on Mn/Y-1222 compound are depicted in Fig. 4. The main panel shows the magnetic susceptibility ($\chi$) versus temperature ($T$) plot for Mn/Y-1222 compound in an applied field of 10 Oe. From 250 K down to around 30 K the compound exhibited normal paramagnetic like behavior, though very small difference could be seen in ZFC and FC at higher temperatures. Both FC and ZFC susceptibility apart substantially from each other at say 30 K. The FC part shoots up very sharply with a ferromagnetic (FM) –like transition. The FM-like nature is further clear from the fact that at below 10 K the FC signal tends slightly towards the typical saturation, though is not fully attained. As far as ZFC is concerned the same exhibited some sort of a down turn transition below 30 K, which is typical to the canted ferromagnetism case of ruthenocuprates. Worth mentioning is the fact that, in present case also we have *nullified* the trapped negative moment in the SQUID before the ZFC process measurements. To further elaborate on this point in Inset I we show the ZFC and FC behavior of $\chi(T)$ of Mn/Y-1222 at applied field of 100Oe. It is clear from inset II that the compound exhibits FM-like transition in both ZFC and FC process. It seems that Mn/Y-1222 orders ferromagnetically at 30 K. To further elaborate this, we show in Inset II of Fig. 4, the $M(H)$ loops at 5 K, within low applied fields of up to 3000 Oe. A clear FM-like opening of the loop is seen having coercive field ($H_c$) of around 300 Oe and returning moment ($M_r$) close to 0.025 $\mu_B$. The compound need to be doped sufficiently with mobile carriers to introduce superconductivity in Cu-O planes along with the currently seen FM of Mn-O blocks. Further work in terms of phase purity and introduction of superconductivity in the magnetic Y/Mn-1222 is yet warranted.

Fig. 5 depicts the magnetic susceptibility ($\chi$) versus temperature ($T$) plot for Cr/Y-1222 compound in an applied field of 10 Oe. The absolute value of susceptibility is quite low, indicating towards nearly non-magnetic Cr and Cu in Cr/Y-1222. Like Mn/Y-122 in case of Cr/Y-1222 also, the ZFC and FC parts away slightly and shoots up to further show FM-like transition. The branching/shoot-up temperature for Cr/Y-1222 is though higher (45 K) than as for Mn/Y-1222 (30 K). Again same question arises, whether the superconductivity could be introduced in this weakly magnetic system or not. The $M(H)$ loops at 5 K within low applied fields of up to 3000 Oe of Cr/Y-1222 are shown in the inset of Fig.5. No opening of the loop is visible, and further the magnetization is linear with field, looking like a normal paramagnetic behavior. This shows that it is less likely that Cr/Y-1222 could qualify for the magneto-superconducting compound, because its magnetism is too weak. In comparison the Mn/Y-1222 looks like more as a possible magneto-superconducting cuprate, provided superconductivity could be introduced in Cu-O planes. We stress further more work is required to introduce superconductivity in these newly formed weakly FM systems.

In conclusion, we can safely conclude that two new possible magneto-superconducting compounds viz. Mn/Y-1222 and Cr/Y-1222 compounds are synthesized in near single-phase form by HPHT process and the phase purity of earlier reported Ru/Y-1222 is further improved significantly.

This work is partially supported by INSA-JSPS bilateral exchange visit of Dr. V. P. S. Awana to NIMS Japan. Authors from the NPL appreciate the interest and advice of Professor Vikram Kumar (Director) in the present work.

**FIGURE CAPTIONS**

Fig. 1 X-Ray diffraction patterns of HPHT synthesized $MSr_2Y_{1.5}Ce_{0.5}Cu_2O_{10}$ compounds, with M = Ru, Mn and Cr.

Fig. 2 $\chi(T)$ behavior for the present $RuSr_2Y_{1.5}Ce_{0.5}Cu_2O_{10}$ at $H$ = 10 Oe, Inset-I shows the same for earlier reported (Ref. 12) 1200 $^0$C synthesized HPHT Y/Ru-1222. Inset –II and III exhibit $M(H)$ loops at 5 K for Ru/Y-1222 in low and high fields respectively.

Fig. 3 $M(H)$ plots for the present $RuSr_2Y_{1.5}Ce_{0.5}Cu_2O_{10}$ at various temperatures from 5 – 150 K, inset shows the $H_c(T)$ plot for the same.

Fig. 4 $\chi(T)$ behavior for the present $MnSr_2Y_{1.5}Ce_{0.5}Cu_2O_{10}$ at $H$ = 10 Oe, Inset-I shows the same for in $H$ = 100 Oe. Inset –II and III exhibit $M(H)$ loops at 5 K for Mn/Y-1222 in low and high fields respectively.

Fig. 5 $\chi(T)$ behavior for the $RuSr_2Y_{1.5}Ce_{0.5}Cu_2O_{10}$, Inset shows the $M(H)$ loops for the same at 5 K.

Fig. 1

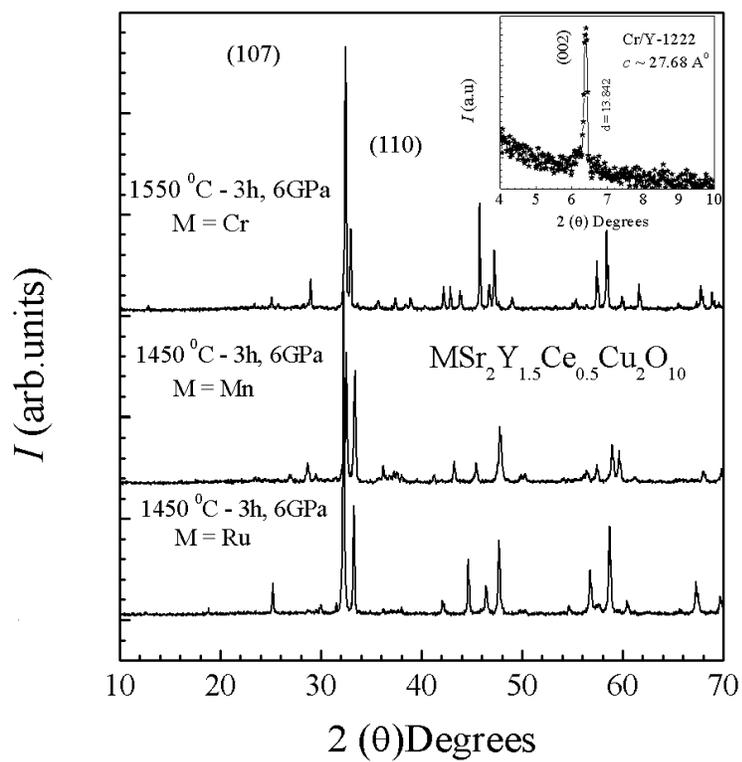

Fig. 2

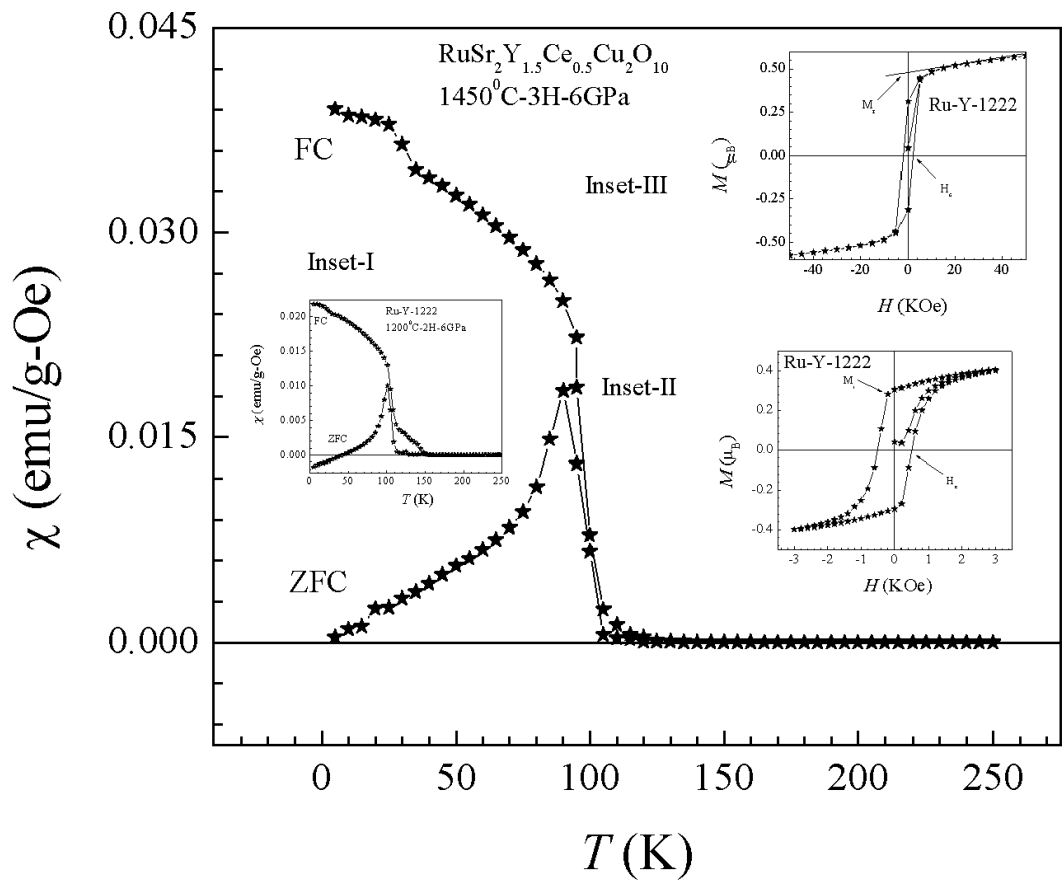

Fig. 3

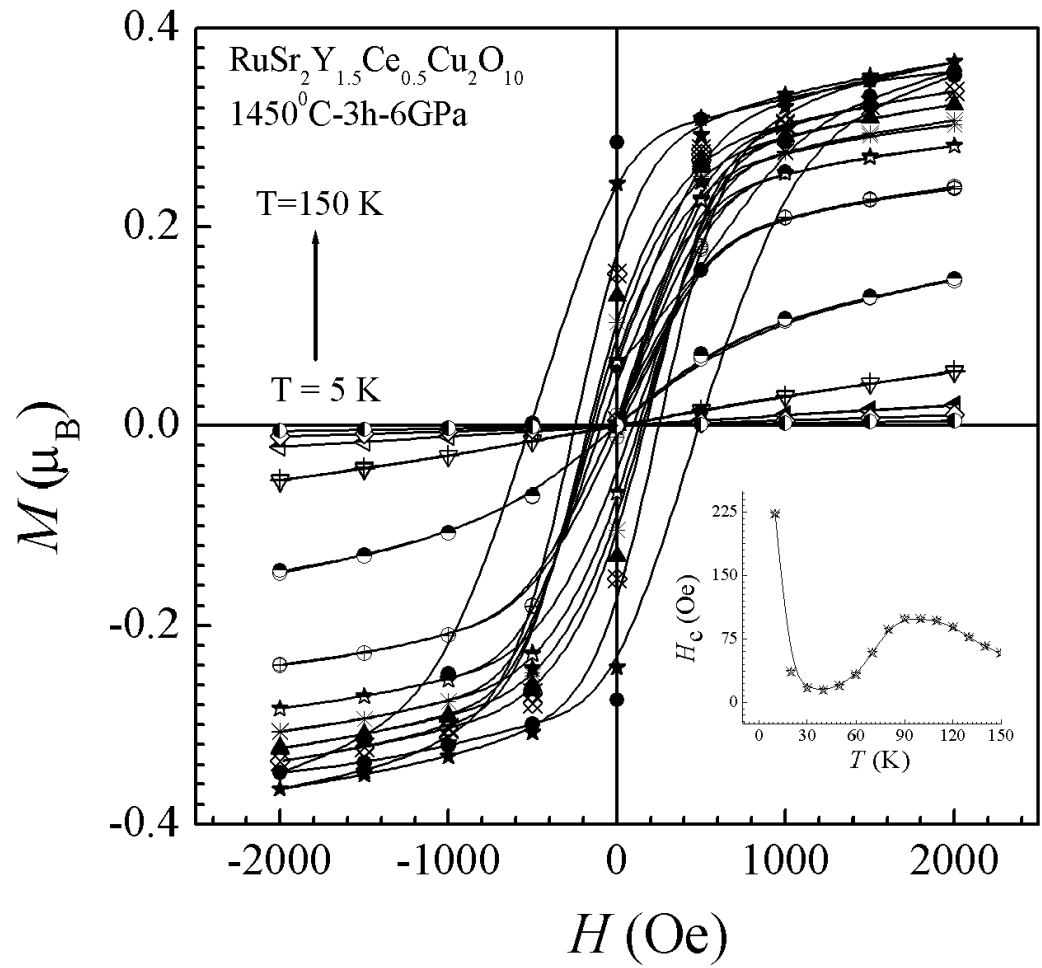

Fig. 4

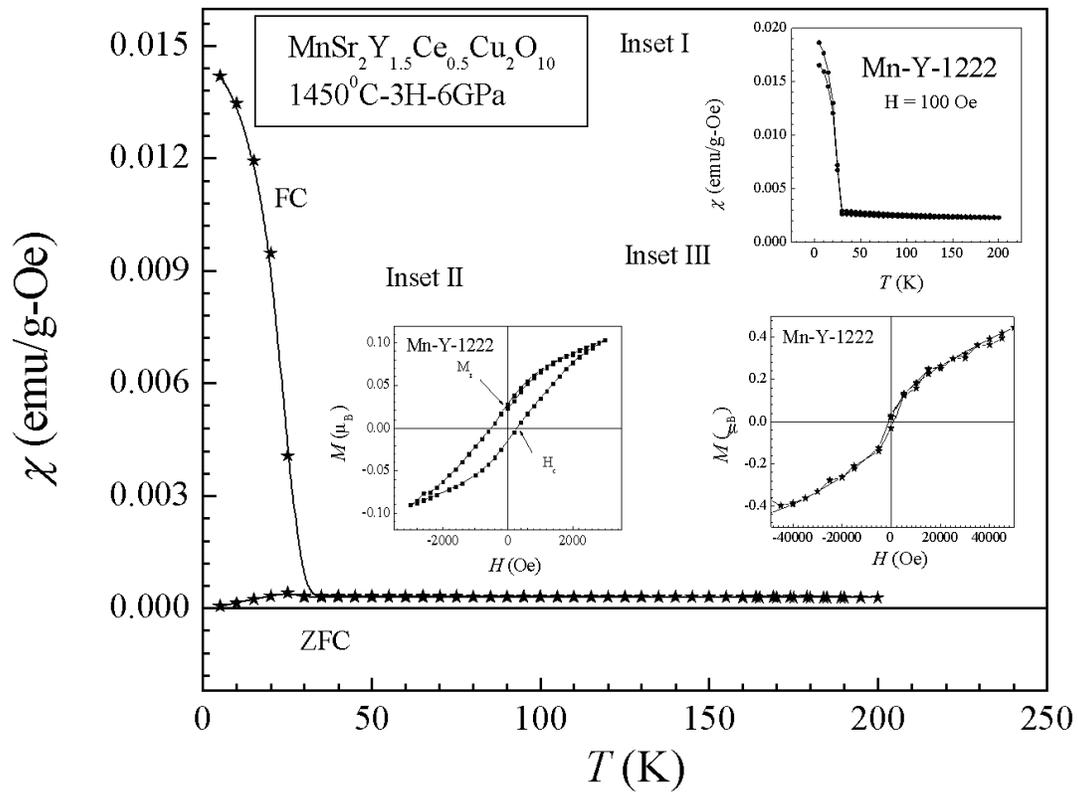

Fig. 5

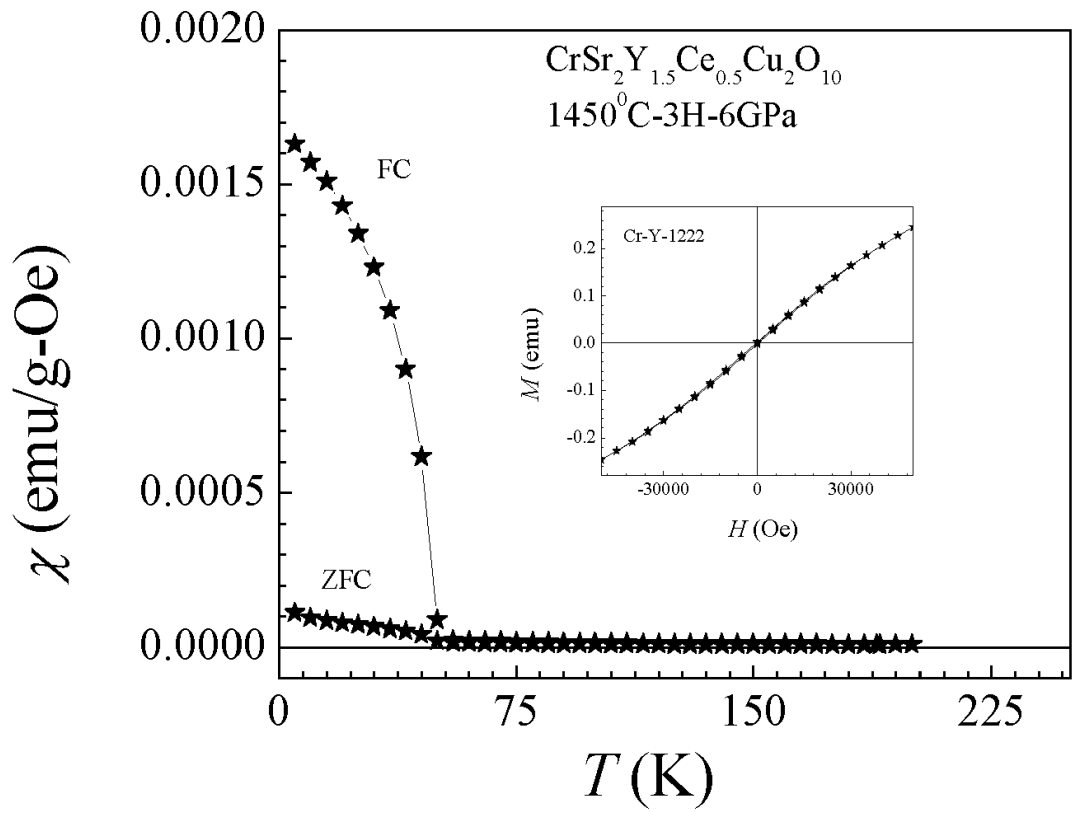